# Physics Education under the Application of Artificial Intelligence: Bibliometric Analysis Based on Web of Science Core Library (2021-2025)

Chengtian Liang[*,#], Yike Qian[#] , Yixuan Lin[#] , Yu Wang[*]

(School of Physics, Hangzhou Normal University, Hangzhou 311121, Zhejiang, China)

*Corresponding Author : lct.lctsoft@hotmail.com (Primary); 20426399@qq.com

#These authors contributes to this article equally.

**Abstract:** The rapid development of artificial intelligence technology is driving the transformation of physics education from traditional models to intelligent and data-driven approaches. To explore the evolution and cutting-edge hotspots in this field, this study conducted a systematic bibliometric analysis of 138 core literature published between 2021 and 2025 using VOSViewer and CiteSpace, based on the Web of Science Core Collection database. Research shows that the number of related publications will increase exponentially from 2023, with the United States, China, and Germany being the main research forces. The research focus has rapidly evolved from early machine learning assisted data analysis to the application of generative AI in teaching, the integration of physics information neural networks in computational physics courses, and the exploration of intelligent medical physics education. At present, this field is in the early stages of explosive growth and exhibits significant interdisciplinary characteristics. Future research should focus on building an adaptive learning ecosystem, reconstructing evaluation systems, and cultivating students' AI ethics and physical intuition.

## 1. Introduction

The rapid development of artificial intelligence technology is currently driving a profound digital transformation in the way physics research is conducted. From traditional theoretical deduction and experimental observation to a new stage centered on data-driven and intelligent computing, artificial intelligence has deeply integrated into all aspects of physical exploration. It can not only assist in discovering unknown physical laws by processing massive experimental data, but also use complex simulation operations to help scientists understand the evolution mechanism of multi-body systems. It can also use deep learning technology for accurate image recognition in astronomical observations or microscopic imaging [1-9]. In addition, more importantly, artificial intelligence is widely used in teaching management, which can generate customized learning paths and experimental guidance plans based on students' learning trajectories and problem-solving records. Related studies have shown [10,11] that compared to other disciplines, introducing artificial intelligence into the field of physics often brings more significant research and teaching returns, and the academic community has gradually begun exploring and practicing the empowerment of physics teaching with artificial intelligence.

Given this, the modernization reform of physics education is imperative.

In recent years, the academic community has conducted a large number of global bibliometric analyses using visual bibliometric software. These works mainly focus on the scientific research output and development trend in cutting-edge fields such as quantum computing [12-14]. However, to the best of the author's knowledge, systematic bibliometric analysis of the current application status of artificial intelligence in physics education is still rare. Therefore, this study aims to use VOSViewer to conduct an in-depth evaluation of the research hotspots and evolutionary trends of artificial intelligence empowering physics teaching over the past five years.

## 2. Research Design

（1） Data source

All literature data involved in this study were collected from the Webof Science core database.The literature search mainly covers and focuses on the two main keywords of "artificial intelligence" and "physics education" (see Table 1 for details), with a time span set from 2021 to 2025. The metadata of this research paper was downloaded from the Web of Science database in. txt format. Through manual screening, 9 papers that did not match the research topic or were of incorrect type were excluded from the initially retrieved 147 papers, and ultimately 138 papers were retained.

（2） Research methods and tools

The analysis method used in this study is bibliometric analysis, and the analysis tool is VOSViewer. VOSViewer is a citation visualization analysis tool developed by Leiden University [15]. This software can present the internal structure, evolution laws, and distribution characteristics of disciplinary knowledge reflected in a series of literature in an intuitive and convenient way. The generated scientific knowledge graph helps researchers quickly identify research hotspots, development trends, and current situations in specific fields. It is a tool focused on bibliometric data processing, adept at constructing graphs that include countries, institutions, authors, journals, and keywords.

**Table 1 Search Query Statements**

| serial number | Number of results | Search Query statement |
|---|---|---|
| #1 | 281371 | (((TS=(Generative AI)) OR TS=(AI)) OR TS=(Artificial Intelligence)) OR TS=(Generative Artificial Intelligence) Indexes=Web of Science Core Collection, timespan=2021-2025 |
| #2 | 2748 | (TS=(physics education))<br>Indexes=Web of Science Core Collection, timespan=2021-2025 |
| #3 | 147 | #1 AND #2 |

## 3. Bibliometrics and Graph Analysis

（1） Annual Publications

The annual publication volume refers to the distribution of literature publication years. The bibliometric analysis of this study ended on November 28, 2025. As shown in Figure 1, in the past 5 years, a total of 138 academic papers have been published exploring the integration of artificial intelligence and physics education. These research results have had a profound impact on promoting physics teaching reform under the application of artificial intelligence.

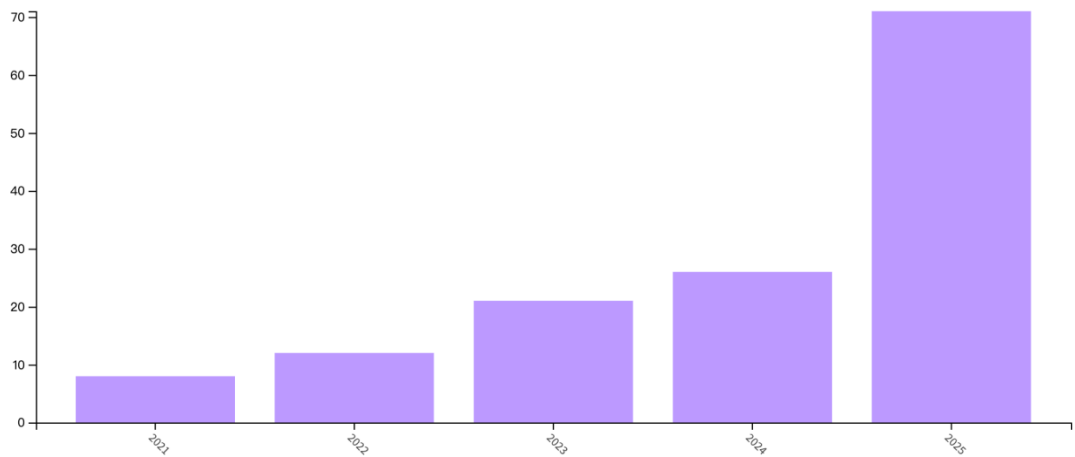

**Figure 1 Annual publication volume chart of relevant literature**

The publication volume of relevant literature showed a sharp increase from 2024 to 2025 and reached a historical peak in 2025. Overall, the flourishing development of artificial intelligence has brought unprecedented opportunities and challenges to the entire physics community. As the cornerstone of the development of basic science, physics education can draw nutrients from the deep application of artificial intelligence, thereby promoting the continuous innovation of the teaching system.

（2） National Analysis

The analysis results are shown in the following chart. Between 2021 and 2025, a total of 62 countries/regions worldwide have participated in the exploration and practice of artificial intelligence in the field of physics education, with 42 countries/regions having at least one cooperative relationship with another country/region. In the context of cooperation with other countries/regions, the United

States ranks first in terms of publication volume with 35 articles, making it the most active research force in this field. Following closely behind are China (16 articles), Germany (16 articles), and Spain (12 articles), with other major contributing countries including the United Kingdom (11 articles), Sweden (9 articles), Indonesia (8 articles), Switzerland (8 articles), Canada (7 articles), and France (7 articles).

Table 2: Top 10 publications, centrality, and citations of countries

| Rank | Number of articles | Countries | Citation | Countries | Total link strength | Countries |
|---|---|---|---|---|---|---|
| 1 | 35 | the United States | 358 | the United States | 62 | France |
| 2 | 16 | China | 320 | Germany | 59 | the United Kingdom |
| 3 | 16 | Germany | 304 | Canada | 54 | Italy |
| 4 | 12 | Spain | 211 | Switzerland | 49 | Netherlands |
| 5 | 11 | the United Kingdom | 164 | Spain | 48 | Spain |
| 6 | 9 | Sweden | 152 | China | 46 | Iceland |
| 7 | 8 | Indonesia | 103 | Sweden | 46 | Germany |
| 8 | 8 | Switzerland | 91 | Italy | 45 | Switzerland |
| 9 | 7 | Canada | 89 | Netherlands | 44 | Greece |
| 10 | 7 | France | 88 | Iceland | 40 | Belgium |

In terms of academic influence, the countries/regions with the highest citation frequency are the United States (358 times), Germany (320 times), and Canada (304 times). In terms of the overall link strength indicator for measuring the closeness of international cooperation, France (62), the United Kingdom (59), and Italy (54) rank in the top three (see Table 2 for details). Figure 2 further reveals that the United States and European countries have shown a clear tendency to integrate artificial intelligence technology into physics teaching, mainly due to their profound technological accumulation and advanced educational infrastructure. Overall, the United States and some European countries have maintained a significant leading position in this field, not only producing a large number of high-level papers, but also establishing extensive and in-depth research cooperation networks with countries around the world.

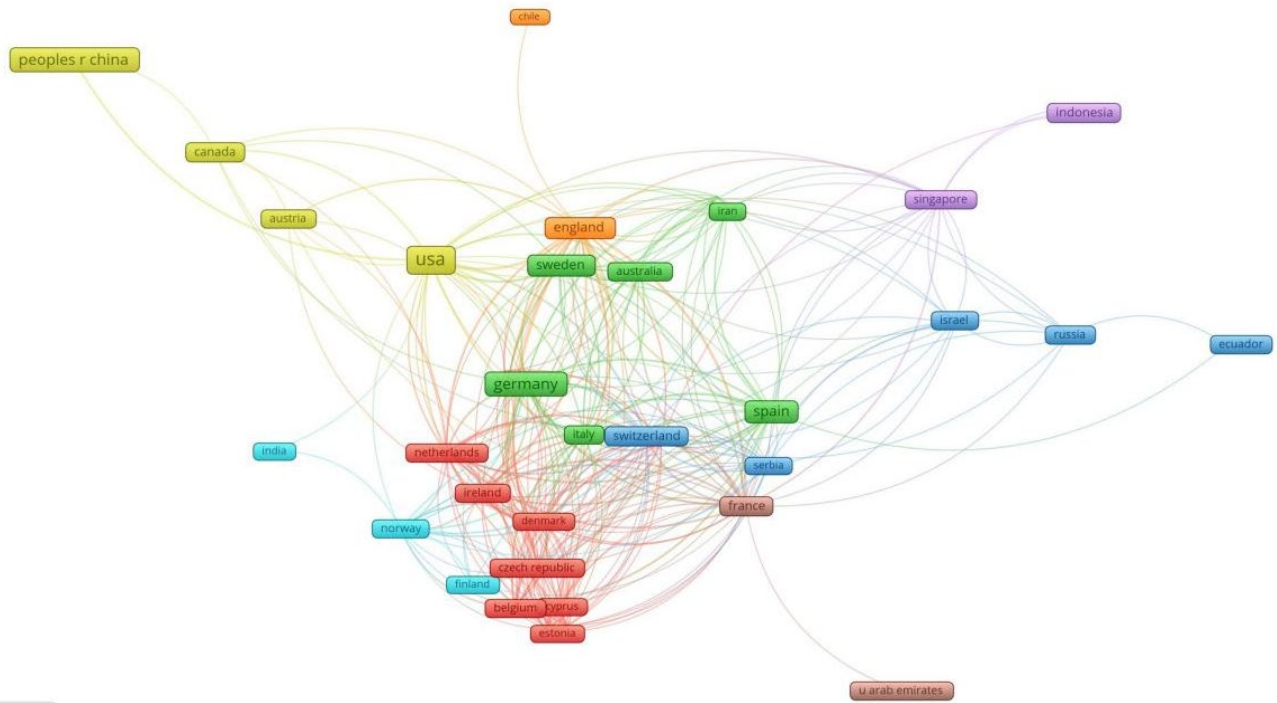

**Figure 2 National and regional co-occurrence map (only showing situations where there is cooperation with at least one other country/region)**

（3） Institutional Analysis

Furthermore, the author will focus on research institutions from an analytical perspective. Data shows that from 2021 to 2025, a total of 338 institutions worldwide have been engaged in research on artificial intelligence in physics education. Among them, 127 research institutions have at least one cooperative relationship with another research institution. In the context of having cooperative relationships with other institutions, the institution ranked first in terms of publication volume is Uppsala University (which has published 5 and 4 papers), followed by Heidelberg University of Education, which has contributed 4 papers (see Table 3 for details).

**Table 3: Top 10 publications, centrality, and citations of institutions**

| Rank | Documents | Institution | Citations | Institution | Total link strength | Institution |
|---|---|---|---|---|---|---|
| 1 | 5 | Uppsala University | 286 | University of Toronto | 41 | National University of Singapore |
| 2 | 4 | Heidelberg University of Education | 132 | Heidelberg University of Education | 41 | Texas A&M University |
| 3 | 3 | Swiss Federal Institute of Technology Zurich | 98 | Swiss Federal Institute of Technology Zurich | 36 | Medical University of South Carolina |
| 4 | 3 | University of Toronto | 93 | North Carolina State University | 35 | Indiana University |
| 5 | 3 | university of Barcelona | 90 | ball state university | 34 | Massachusetts Institute of Technology |
| 6 | 3 | Massachusetts Institute of Technology | 90 | University of Alabama | 34 | Harvard University |
| 7 | 3 | Stanford University | 89 | Uppsala University | 29 | Stockholms universitet |
| 8 | 3 | Purdue University | 89 | Michigan state university | 26 | Stanford University |
| 9 | 3 | Texas A&M University | 71 | university of Barcelona | 26 | University College London |
| 10 | 3 | Taiwan Normal University | 70 | Leiden University | 24 | Purdue University |

In terms of academic influence, the University of Toronto (286 citations), Heidelberg University School of Education (132 citations), and Swiss Federal Institute of Technology Zurich (98 citations) rank in the top three. This indicates that the research achievements of these institutions in the field of artificial intelligence in physics education have received widespread attention in the industry. In addition, the collaborative network analysis in Figure 3 further reveals that the National University of Singapore and Texas A&M University are the most active in cross institutional collaboration, with a total link strength of 41 for both, demonstrating their core hub position in building research collaboration networks.

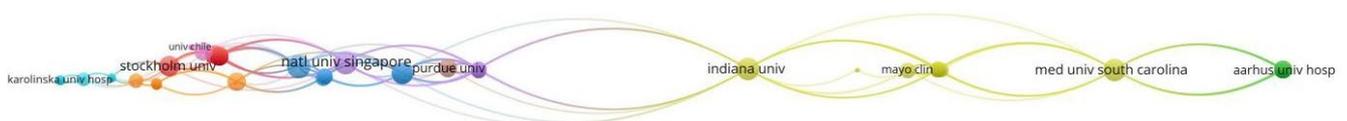

**Figure 3 Institutions co-occurrence map (only showing situations where there is at least one cooperative relationship with other institutions)**

（4）Author Analysis

Statistics show that in the past five years, a total of 662 scholars have devoted themselves to the research of the integration of artificial intelligence and physics education. Among them, 64 scholars have a cooperative relationship with at least one other scholar. In the context of collaborative relationships with other scholars, statistical analysis of their academic output is conducted, Bor Gregorcic, Giulia Polverini, Peter Wulff, Marcus Kubsch, Ralf Widenhorn, Gerd Kortemeyer Scholars are the most active, with each contributing more than 3 papers. It is worth noting that among these prolific authors, only Peter Wulff and Gerd Kortemeyer have also taken the top 5 citation rankings, with each author's cumulative citation count not less than 98 times (see Table 4 for details).

Table 4: Top 10 publications, centrality, and citations of authors

| Rank | Documents | Author | Citations | Author | Total link strength | Author |
|---|---|---|---|---|---|---|
| 1 | 5 | Bor Gregorcic | 282 | Rajesh Bhayana | 63 | Andrey Ustyuzhanin |
| 2 | 5 | Giulia Polverini | 282 | Robert R. Bleakney | 62 | Keir Adams |
| 3 | 4 | Peter Wulff | 282 | Satheesh Krishna | 62 | Tianfan Fu |
| 4 | 3 | Marcus Kubsch | 132 | Peter Wulff | 62 | Nicolas Gao |
| 5 | 3 | Ralf Widenhorn | 98 | Gerd Kortemeyer | 62 | Erik Bekkers |
| 6 | 3 | Gerd Kortemeyer | 90 | Lu Ding | 62 | Michael Bronstein |
| 7 | 2 | Dazhen Tong | 90 | Albert Gapud | 62 | Carl Edwards |
| 8 | 2 | Christoph Bert | 90 | Shiyan Jiang | 62 | Stefano Ermon |
| 9 | 2 | Zeyu Zhang | 90 | Tong Li | 62 | Stephan Gunnemann |
| 10 | 2 | Hangxin Liu | 89 | Bor Gregorcic | 62 | Jacob Helwig |

However, from the co-occurrence map generated by VOSViewer, it can be observed that there is currently no core author with absolute dominance in this field, and the number of personal publications is generally low. This phenomenon reflects that the overall research volume in this field is still at a preliminary level. As shown in Figure 4, the research on the application of artificial intelligence in physics education is still in its early exploratory stage, but there is a trend of forming a monopolistic research team with significant leading advantages, and the composition of the authors of this monopolistic collaborative network has undergone significant changes between 2023 and 2025.

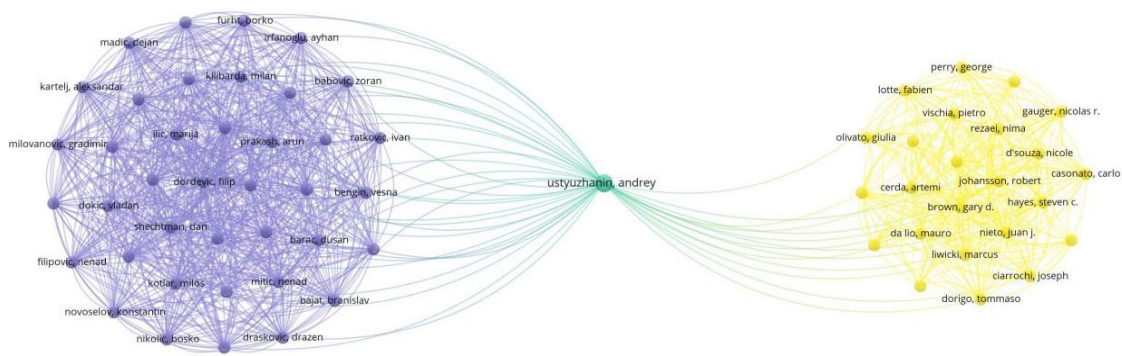

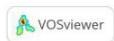
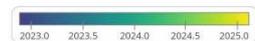

**Figure 4 Authors' co-occurrence map (only showing cases where there is at least one cooperative relationship with other institutions)**

（5）References Analysis

As shown in Table 5, we have selected 15 representative key literature, which constitute the core reference materials for current research on the integration of artificial intelligence and physics education. Among them, the paper titled "Could an artificial intelligence agent pass an introductory physics course?" performed the most outstandingly, and it is currently a research result with strong citation frequency (85 times) and literature relevance (23 times). This article evaluates the performance of ChatGPT in basic physics courses and finds that although it can barely pass, it often makes beginner's mistakes. Based on this, it explores the potential impact of AI on physics teaching, assessment, and research.

**Table 5: Top 10 publications, centrality, and citations of cited reference**

| Rank | Citations | Cited reference, year | Total link strength | Cited reference, year |
|---|---|---|---|---|
| 1 | 282 | Bhayana (2023)[16] | 23 | Kortemeyer (2023)[18] |
| 2 | 90 | Ding (2023)[17] | 16 | Polverini (2024b)[26] |
| 3 | 85 | Kortemeyer (2023)[18] | 15 | Dahlkemper (2023)[23] |
| 4 | 69 | Huang (2023)[19] | 15 | Polverini (2024a)[24] |
| 5 | 69 | Li (2022)[20] | 14 | Kieser (2023)[22] |
| 6 | 61 | Garcia Martinez (2023)[21] | 14 | Kortemeyer (2025)[27] |
| 7 | 55 | Kieser (2023)[22] | 11 | Polverini (2025b)[28] |
| 8 | 53 | Dahlkemper (2023)[23] | 9 | Kortemeyer (2024)[29] |
| 9 | 49 | Polverini (2024a)[24] | 8 | Ding (2023)[17] |
| 10 | 43 | Liang (2023)[25] | 8 | Wulff (2024)[30] |

Only in terms of the number of arguments, the top ranked ones are "Performance of ChatGPT on a Radiology Board style Examination: Insights into Current Strengths and Limitations" and "Students' perceptions of using ChatGPT in a physics class as a virtual tutor". Figure 5 visually illustrates the distribution pattern of the core literature mentioned above and their interrelationships in academic networks.

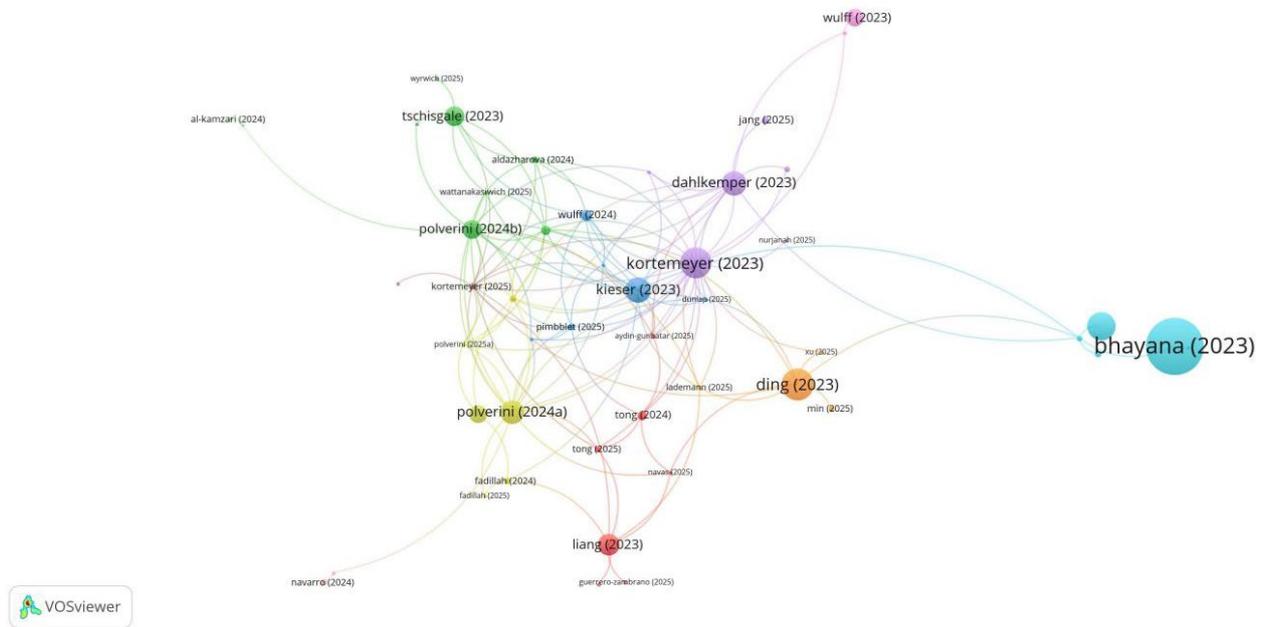

**Figure 5 Cited reference co-occurrence map (only showing cases where there is a relevant relationship with at least one other reference)**

（6）Keyword situation

The research on the integration of artificial intelligence and physics education from 2021 to 2025 focuses on 12 main keywords (Table6). Figure 6 shows that artificial intelligence (38), education (22), chatgpt (20), and machine learning (12) have the highest frequency and correlation strength.

**Table 5: Top 10 Keywords Related to AI in Physics Education**

| Rank | Occurrence (%) | Keywords | Total link strength | Keywords |
|---|---|---|---|---|
| 1 | 38 | artificial intelligence | 298 | artificial intelligence |
| 2 | 22 | education | 163 | education |
| 3 | 20 | chatgpt | 122 | chatgpt |
| 4 | 12 | machine learning | 117 | machine learning |
| 5 | 12 | physics education | 88 | physics |
| 6 | 10 | physics | 72 | artificial-intelligence |
| 7 | 10 | artificial-intelligence | 71 | natural language processing |
| 8 | 8 | ai | 66 | physics education |
| 9 | 8 | generative artificial intelligence | 61 | impact |
| 10 | 7 | natural language processing | 60 | large language models |

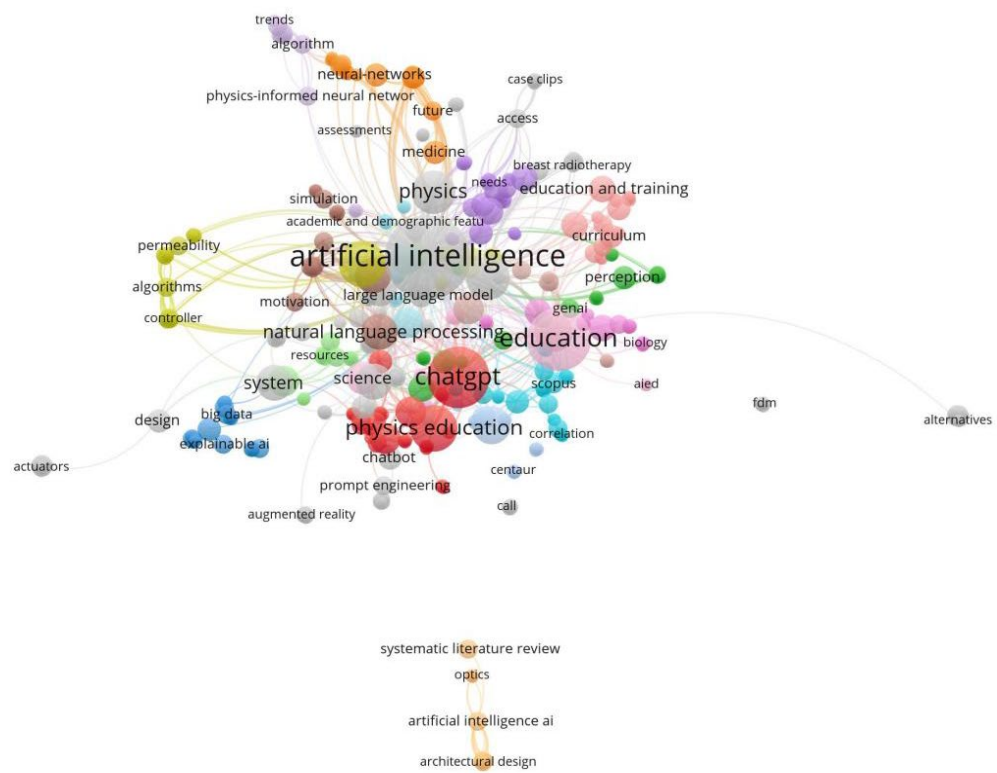

**Figure 6 Keywords co-occurrence map (only showing cases where there is a relevant relationship with at least one other keyword)**

## 4. Results and Detailed Explanation

As shown in Figure 6 (keyword co-occurrence map) and Figure 5 (reference co-occurrence map), through clustering analysis of highly cited literature and core keywords, we can identify the highly acclaimed research achievements and core themes of artificial intelligence in the field of physics education. The research hotspots between 2021 and 2025 mainly focus on the following five core clusters, which reflect how artificial intelligence technology gradually permeates and reshapes the teaching and learning of physics.

Cluster 1: Generative Artificial Intelligence and the Reconstruction of Physics Teaching Paradigm

In the central region of the keyword co-occurrence network in Figure 6, "Artificial Intelligence", "ChatGPT", "Generative AI", and "Large Language Models" constitute the core clusters with the highest connectivity. This is highly consistent with the trend of a significant increase in the number of publications after 2023 shown in Figure 1, indicating that generative big language models represented by ChatGPT have become the absolute focus of current research on the integration of physics education and artificial intelligence.

The research on this clustering mainly focuses on evaluating AI's dual role ability as a "physics learner" and "physics mentor". As shown by the highly cited literature Bhayana (2023) (282 citations) and Kortemeyer (2023) (85 citations) ranked first and third in Table 5, the academic community's first focus is on the performance of AI models in standardized exams. It is worth noting that the top ranked Bhayana (2023), although a study on radiology exams, is widely cited in the physics education community as an effective model for evaluating AI exam abilities, to compare the cognitive level of AI in different disciplines. Kortemeyer (2023) specifically evaluated the performance of ChatGPT in introductory physics courses and found that although AI can pass the exam, it still exhibits "pre concept" errors similar to beginners when dealing with complex physics concepts.

The adjacent "Prompt Engineering" and "Chatbot" nodes in the graph indicate that educators are shifting from the initial "technology panic" to "technology integration", exploring how to use carefully designed prompt words to assist in the development of physics teaching resources.

Cluster 2: Physical Information Neural Networks and Computational Physics Education

This is a key cluster that was easily overlooked in the previous analysis, but must be emphasized and supplemented according to Figure 6. In the upper yellow area of Figure 6, nodes such as "Neural Networks", "Physics informed neural networks", and "Algorithm" are clearly visible.

This discovery indicates that AI research in physics education is not only about using AI as a teaching aid, but also includes the content of "teaching AI to solve physics problems". Physical Information Neural Networks (PINNs), as a deep learning method that embeds physical laws (such as partial differential equations) into neural network loss functions, are becoming a new frontier in computational physics education. The current research status shows that some physics courses have begun to introduce PINNs, teaching students how to use AI algorithms to solve complex equations in fluid mechanics or quantum mechanics [7,31]. This marks the transition of physics education from traditional numerical simulations (such as finite element analysis) to AI driven scientific computing, with AI itself becoming a part of the physics curriculum.

Cluster 3: Medical Physics Education and Interdisciplinary Applications

Based on the purple clustering area on the right side of Figure 6, we found significant nodes such as "Medical Physics", "Medicine", "Breast Radiotherapy", and "Education and training".

The existence of this clustering reveals an important branch in physics education research - medical physics education. Due to the high dependence of medical physics on image processing and dose calculation, the intervention of AI technology is particularly natural and urgent. The current research status indicates that AI has been widely used to assist teaching in graduate education and vocational training in medical physics, especially in the development of radiation therapy plans and analysis of medical images [32,33]. Educational research in this field often has strong career orientation, focusing on cultivating students' ability to use AI tools in clinical environments, and is a typical representative of the cross integration of physics and medical education.

Cluster 4: Machine Learning Driven Data Mining and System Construction

The blue node cluster in the lower left corner of Figure 6, including "Big Data", "Machine Learning", "System", and "Explainable AI", constitutes the data-driven clustering of physics education research.

The research in this field focuses on using machine learning algorithms to mine massive amounts of data in the process of physics education. For example, by analyzing students' answer logs and click data on online physics homework platforms, machine learning models can accurately predict students' academic performance. The appearance of the "Explainable AI" node in the graph is also worth our attention. This indicates that researchers are not only concerned with the predictive accuracy of algorithms, but also with the interpretability of algorithms, attempting to explore which specific physical concepts or mathematical skills are the key factors causing students' learning difficulties, in order to provide theoretical basis for the construction of personalized teaching systems.

Cluster 5: Teacher Student Cognition and Curriculum Evaluation

In the green area on the right side of the graph, nodes such as "Curriculum", "Perception", "Students", and "Needs" form research clusters about "people".

Based on Ding (2023) (90 citations) and other literature in Table 5, this clustering focuses on the attitudes of entities in the physics education ecosystem towards technology. The current research indicates that the attitudes of physics students and teachers towards AI exhibit complex duality [17,34-36]. On the one hand, students generally welcome AI as a 24/7 virtual mentor; On the other hand, the negative impact of AI ethics, academic integrity, and the "AI illusion" on the understanding of physical concepts constitute the core topics of current curriculum reform discussions.

**5. Summary and Prospect**

This study utilized the VOSViewer tool to conduct a visual quantitative analysis of 138 core literature in the field of "Physics Education under Artificial Intelligence Applications" from 2021 to 2025. The research results indicate that the field is in the early stages of explosive growth, especially since 2023, where technological breakthroughs represented by big language models have greatly accelerated the output of related research. From a spatial distribution perspective, the United States, China, and European countries constitute the main research landscape, and have formed several cross institutional cooperation networks with international influence.

From the evolution of research content, the application of AI in physics education has shown obvious hierarchical characteristics: firstly, reconstructing teaching interaction and evaluation paradigms with generative AI; secondly, promoting the modernization of computational physics courses with physics information neural networks; thirdly, vocational application education represented by medical physics. This indicates that AI is no longer just an external tool for assisting teaching, but is being internalized as a part of physics teaching content.

In the future, the deep integration of artificial intelligence and physics education will face a new normal of "opportunities and challenges coexisting". On the one hand, physics teachers need to use AI to build a truly adaptive learning ecosystem and solve personalized teaching problems; On the other hand, how to deal with the interference of AI "illusions" on the learning of physical concepts, how to redefine the "core physical literacy" of the intelligent era, and how to establish academic ethical boundaries for human-machine collaboration will be key issues that must be overcome in future research. The ultimate goal of physics education will shift from "cultivating people who are good at computation" to "cultivating scientists who are good at mastering artificial intelligence to explore the laws of the universe".